\documentclass[conference]{IEEEtran}
\IEEEoverridecommandlockouts

\usepackage{cite}
\usepackage{amsmath,amssymb,amsfonts}
\usepackage{amsthm}
\usepackage{algorithmic}

\usepackage{graphicx}
\usepackage{textcomp}
\usepackage{xcolor}
\usepackage[normalem]{ulem}
\def\BibTeX{{\rm B\kern-.05em{\sc i\kern-.025em b}\kern-.08em
    T\kern-.1667em\lower.7ex\hbox{E}\kern-.125emX}}

\usepackage{mathtools}

\newtheorem{theorem}{Theorem}[section]

\newtheorem{lemma}[theorem]{Lemma}
\usepackage{algorithm}

\usepackage{subcaption}
\begin{document}

\title{How Long Can I Transmit? A Mobility Aware mmWave-based UAV Communication Framework}

\author{
    \IEEEauthorblockN{Shawon Mitra\IEEEauthorrefmark{1}, Subhojit Sarkar\IEEEauthorrefmark{2}, Sasthi C. Ghosh\IEEEauthorrefmark{1}}
    \IEEEauthorblockA{\IEEEauthorrefmark{1}Advanced Computing and Microelectronics Unit, Indian Statistical Institute, Kolkata, India\\Email : shawon.mitra@gmail.com, sasthi@isical.ac.in}
    \IEEEauthorblockA{\IEEEauthorrefmark{2}Indian Institute of Technology, Goa, India\\Email : subhojit@iitgoa.ac.in}
}

\maketitle

\begin{abstract}
One primary focus of next generation wireless communication networks is the millimeterwave (mmWave) spectrum, typically considered in the 30 GHz to 300 GHz frequency range. Despite their promise of high data rates, mmWaves suffer from severe attenuation while passing through obstacles. Unmanned aerial vehicles (UAVs) have been proposed to offset this limitation on account of their additional degrees of freedom, which can be leveraged to provide line of sight (LoS) transmission paths. While some prior works have proposed analytical frameworks to compute the LoS probability for static ground users and a UAV, the same is lacking for mobile users on the ground. In this paper, we consider the popular Manhattan point line process (MPLP) to model an urban environment, within which a ground user moves with a known velocity for a small time interval along the roads. We derive an expression for the expected duration of LoS between a static UAV in the air and a mobile ground user, and validate the same through simulations. To demonstrate the efficacy of the proposed analysis, we propose a simple user association algorithm that greedily assigns the UAVs to users with the highest expected LoS time, and show that it outperforms the existing benchmark schemes that assign the users to the nearest UAVs with LoS without considering the user mobility.
\end{abstract}

\section{Introduction}
    \noindent Though primarily developed with the military in mind, civilian utilities of unmanned aerial vehicles (UAVs) have seen a sharp rise in recent years. From aerial mapping to smart agriculture, from package delivery to drone shows, the opportunities are limitless. As the world becomes more and more connected, UAVs are poised to be an important enabling technology for next generation communication networks. Typical use cases of UAVs in the wireless communication scenario involve increasing the coverage area, relaying, and collection of delay tolerant information from low power sensing devices. Among these, of particular interest is the first one, providing coverage to terrestrial users from the air. UAVs can come to the rescue if the terrestrial infrastructure becomes incapable of providing service, either due to damage \cite{9696188} or sudden spatial spike in user demand \cite{8329013}. Their fast, on-demand deployability makes them ideal candidates for such applications.
    
    Spatial flexibility offered by these UAVs are of special interest in millimeterwave (mmWave) communications \cite{8641426}.  Millimeterwave communications have seen tremendous interest over the last decade, fueled by their promise of large bandwidth, and communication speeds in the range of gigabits per second \cite{6515173}. The increase in bandwidth hungry applications like full HD streaming, and virtual reality has led to further interest in the field. However, in spite of the high bandwidth, and the subsequent high speeds provided by mmWaves, they have some inherent drawbacks. Communication over these low wavelength signals are extremely prone to blockages \cite{8373698}. This problem is especially exacerbated in urban scenarios, where the dense distribution of buildings causes a significant number of links to be blocked. Measurements show that the penetration loss caused due to a building can be as high as 40dB \cite{6515173}, which causes noticable degradation in quality of service. Consequently, mmWave can provide high data rates only if the UAV is in line of sight (LoS) from the user \cite{8373698}. It is no wonder thus, that computation of the LoS probability between the user and the UAV is an important problem.

    Considerable work has been done to capture the LoS probability in wireless networks,  most of which leverages stochastic geometry. The benefit of stochastic models when compared to ray tracing methods like \cite{7592474} is that they are computationally cheaper, and produces faster network evaluation with acceptable accuracy \cite{8370679}. A few models have been proposed by 3GPP\cite{zhu20213gpp} and ITU \cite{ITU} to consider the LoS probability of wireless communication systems deployed in urban scenarios. A Poisson point process (PPP) was used in \cite{6042301} to model the locations of terrestrial base stations, and subsequently provide a tractable closed form expression on coverage probability. A similar coverage analysis was done for mmWave networks in \cite{6932503}, where the base stations were considered to form a PPP with a fixed density. While the previous works dealt with 2D distributions of base stations, \cite{8580601} additionally considered the heights of the transmitters and receivers to be drawn from the exponential distribution, to compute the blockage effects in 3D. Using concepts from random shape theory, authors in \cite{6840343} proposed an analytical framework for LoS probability for an environment with buildings deployed irregularly. Regular urban grids have also been considered while computing the LoS probability. Authors in \cite{7218450} modeled an urban grid, with roads and buildings generated using the Manhattan Poisson line process (MPLP) featuring two homogenous PPPs. Closed form LoS probability expressions were provided by authors in \cite{9419751} considering different building height distributions. A similar study was done in \cite{9653132}, where an urban tactical mmWave network was considered with a fleet of UAVs, and a vehicle on the ground. Leveraging tools from stochastic geometry, a closed form expression for the LoS connectivity probability was presented.

    One noticeable thing in the previous works is the absence of user mobility, \emph{i.e.}, they consider the ground user to be static during a small time epoch, while computing the LoS probability. As a result, they disregard the fact that a user experiencing LoS with the UAV at the start of a time interval does not always have an LoS throughout the entire duration of the interval. In an urban environment, the presence of buildings will cause intermittent blockages of the mmWave links, thereby leading to network inefficiencies. In this paper, we propose an analytical method to model the expected time under LoS for a mobile ground user connected with a UAV. More precisely, we consider that the user is moving with a known velocity for a short time interval, while communicating with a UAV. In the urban grid environment, we consider expressions for instantaneous LoS probability between the two devices, one on ground and one in the air. We thereafter compute the expectation of the time under LoS from first principles. Two cases arise, for one of which we find a closed form expression, while for the other, we use Simpson's 1/3 rule to get an approximate solution. We finally incorporate the two together to propose an analytical expression that gives us the expected LoS time a moving user has with a static UAV in an urban environment. To show the utility of this metric, we propose a greedy algorithm to associate mobile users with UAVs, with an aim to increase their expected LoS time, and show the superiority of our approach compared to an existing benchmark scheme as commonly used in \cite{10499959,10064007,10198895} that greedily assigns the user to the nearest UAV providing LoS service. To the best of our knowledge, this is the first work that attempts to model the expected time under LoS for a mobile ground user using tools from stochastic geometry, and proposes a user association algorithm based on it. 
    
    The rest of the paper is arranged as follows. Section \ref{sysm} details the system model, while the analysis and the user association are done in Section \ref{elos}. The proposed models are validated through simulation and compared with the existing works in Section \ref{simu}. We conclude in Section \ref{concl}.

\section{System Model}\label{sysm}
\begin{figure}
     \centering
     \begin{subfigure}{0.2\textwidth}
         \centering
         \includegraphics[width=\linewidth]{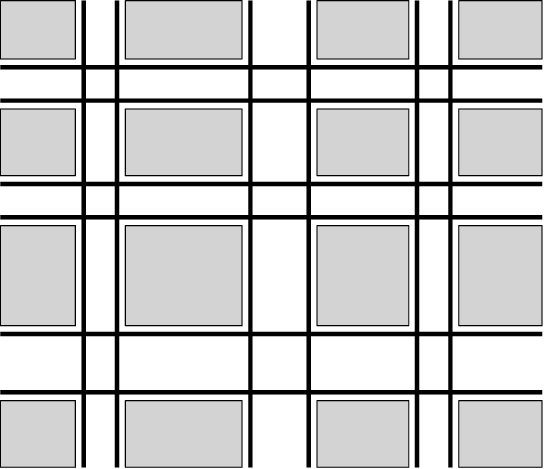}
         \caption{Top view}
         \label{fig:1a}
     \end{subfigure}
     \begin{subfigure}{0.25\textwidth}
         \centering
         \includegraphics[width=\linewidth]{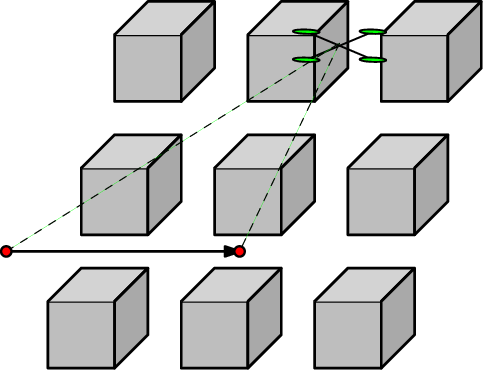}
         \caption{Side view: LoS and Non-LoS}
         \label{fig:1b}
     \end{subfigure}
     \caption{Considered urban environment}
     \label{fig:1}
     \vspace{-15pt}
\end{figure}

\noindent We consider the following urban layout model as in \cite{9419751,9653132}. Roads run orthogonally, either along the X-axis from east to west, or along the Y-axis from north to south. The starting locations of the buildings are drawn from an MPLP, which basically comprises two independent homogeneous PPPs $X$ and $Y$ each with density $\lambda$. The region between two consecutive points on an axis contains a road, and a building block as shown in Fig. \ref{fig:1a}. The mean street and building widths are considered to be $\mu_s$ and $\mu_b$ respectively, while $\lambda=\frac{1}{\mu_s+\mu_b}$.  Along each axis, the ratio of building widths is $\frac{\mu_b}{\mu_b+\mu_s}$, while the ratio of street widths is $\frac{\mu_s}{\mu_b+\mu_s}$ with respect to the total length. 

It is assumed that each non-road section comprises a single building, whose height is a random variable $H_B$ and $F_{H_B}(h)$ is the cumulative distribution function (CDF) of the building height. Since we are dealing entirely with communications in the mmWave range, we consider that building walls are thick enough to cause sufficient signal attenuation to cause link failure. In other words, anything other than optical LoS is considered as a failed path. As in Fig. \ref{fig:1b}, the 3D path is occluded only when 1) the 2D projection of the transmission path crosses a building side, and 2) the height of the said building is greater than the height of the transmission path at the point of 2D intersection. For analysis, we assume without loss of generality that the user is at ground level, and that a single UAV is flying at fixed 3D coordinates. The user is moving with a fixed velocity $v$ along a road, which without loss of generality can be considered to be along the positive X-axis. The user moves with the chosen velocity with no change in direction for a small time interval $[0,T]$. The case of direction change during the time interval $[0,T]$ will be considered in a subsequent future work, and is thus out of scope of the current paper.

\subsection{LoS Probability for Static User}
\noindent Here assuming that the user remains static during the time interval $[0,T]$, we consider the LoS probability between the user and the UAV. The positions of the user and the UAV are $\hat{g_0} = (x_0,y_0,0)$, $\hat{u} = (x_u, y_u,h_u)$, respectively, and the width of the street containing the user is  $w$. Without loss of generality, we can consider the UAV to be at either first or second quadrants. Based on this scenario, the following formula (\ref{basic p-los}) can provide the probability of the LoS between the user and the UAV, as in \cite{9419751},
\begin{align}
  P_{LoS}(\hat{g_0},\hat{u})=P^0_{LoS}(\hat{g_0},\hat{u}).P^X_{LoS}(\hat{g_0},\hat{u}). P^Y_{LoS}(\hat{g_0},\hat{u})
  \label{basic p-los}
\end{align}

\noindent $P^0_{LoS}$ is the probability that the height of the nearest contact building along the projected LoS link is lower than the LoS height, $h_1(\hat{c},\hat{g_0},\hat{u})$, at the point of intersection $\hat{c}=(\tilde{x},\tilde{y})$, $\emph{i.e.}$,
\begin{align}
    P^0_{LoS} = P(H_B\leq h_1(\hat{c},\hat{g_0},\hat{u})) = F_{H_B}(h_1(\hat{c},\hat{g_0},\hat{u}))
    \label{p-los0}
    \vspace{-0.2cm}
\end{align}
where $H_B$ is the height of the nearest contact building and
\begin{align}
 h_1(\hat{c},\hat{g_0},\hat{u})= \frac{h_u|\tilde{x}-x_0|}{|x_u-x_0|} = \frac{h_u|\tilde{y}-y_0|}{|y_u-y_0|}. \label{h1}   
\end{align}
\noindent $P^X_{LoS}$, $P^Y_{LoS}$ are the void probabilities of two nonhomogeneous PPPs along the X and Y axes in the interval $[\tilde{x},x_u]$ and $[\tilde{y},y_u]$ respectively, i.e.,

\begin{align}
  P^X_{LoS} = \exp\Big(-\lambda\int_{\tilde{x}}^{x_u}[1-F_{H_B}(h^x(x,\hat{g_0},\hat{u}))]dx\Big)\label{p-x}\\
  P^Y_{LoS} = \exp\Big(-\lambda\int_{\tilde{y}}^{y_u}[1-F_{H_B}(h^y(y,\hat{g_0},\hat{u}))]dy\Big)\label{p-y}
\end{align}
where $h^x$ and $h^y$ are the LoS heights at $(x,0)$ and $(0,y)$ respectively and
\begin{align}
    h^x(x,\hat{g_0},\hat{u}) = \frac{h_u|x-x_0|}{|x_u-x_0|}, \;\; h^y(y,\hat{g_0},\hat{u}) = \frac{h_u|y-y_0|}{|y_u-y_0|}.
    \label{height}
\end{align}

Now, let the height of the building follow Rayleigh distribution 
with parameter $\sigma$ 
($H_B \sim Rayleigh(\sigma)$), the equations (\ref{p-los0}), (\ref{p-x}), (\ref{p-y}) will be,
\begin{align}
P^0_{LoS} &= \left[ 1 - \exp\left( -\frac{h_1(\hat{x}_0,\hat{u})^2}{2\sigma^2} \right) \right]
\label{eq:p-los0}\\
P^X_{LoS} &= \exp\Big(-\lambda\int_{\tilde{x}}^{x_u}\exp\left( -\frac{h^x(x,\hat{x}_0,\hat{u})^2}{2\sigma^2} \right)dx\Big)
\label{eq:p-losx}\\
P^Y_{LoS} &= \exp\Big(-\lambda\int_{\tilde{y}}^{y_u}\exp\left( -\frac{h^y(y,\hat{x}_0,\hat{u} )^2}{2\sigma^2} \right)dy\Big)
\label{eq:p-losy}
\end{align}
Using (\ref{h1}), (\ref{height}), (\ref{eq:p-los0}), (\ref{eq:p-losx}) and (\ref{eq:p-losy}),  we can rewrite \eqref{basic p-los} as
\begin{align}
     P_{LoS}(\hat{g_0},\hat{u}) =& \Big[1- exp\Big(-\frac{h_u^2}{2\sigma^2}\frac{{|\tilde{x}-x_0|}^2}{{|x_u-x_0|}^2}\Big)\Big]\notag\\\times &\exp\Big( A|x_u - x_0|+ B|y_u - y_0| \Big)
     \label{eq:p_los}
\end{align}

where $A$ and $B$ have the following expressions
\begin{align}
    A = -\lambda\sqrt{\frac{\pi}{2}}\frac{\sigma}{h_u}(\operatorname{erf}(\frac{h_u}{\sqrt{2}\sigma}) - \operatorname{erf}(\frac{h_u}{\sqrt{2}\sigma} \frac{| \tilde{x}-x_0|}{|x_u - x_0|}))\nonumber
\end{align}
\begin{align}
    B = -\lambda\sqrt{\frac{\pi}{2}}\frac{\sigma}{h_u}(\operatorname{erf}(\frac{h_u}{\sqrt{2}\sigma}) - \operatorname{erf}(\frac{h_u}{\sqrt{2}\sigma}  \frac{| \tilde{y}-y_0|}{|y_u - y_0|}))\nonumber
\end{align}
where $\operatorname{erf}(z) = \frac{2}{\sqrt{\pi}}\int_0^z\exp(-t^2)dt$.
\section{Mathematical Analysis for Mobile User}\label{elos}
\noindent In this section, we consider the scenario where the user is not static. Since we consider an urban layout, the buildings will occlude allocated links, thereby causing intermittent link failures due to user mobility. First, in subsection \ref{elostime}, we derive an expression of the expected LoS time of the user with a given UAV for a given time epoch $[0,T]$, based on our system model. Here we assume that the time epoch $T$ is small, so that the user does not change its direction and the velocity $v$ during $[0,T]$. Thereafter, in subsection \ref{uavasso}, we derive a user association policy where a user is associated with the UAV providing the highest expected LoS time.

\subsection{Expected LoS Time}\label{elostime}
\begin{figure}
     \centering
     \begin{subfigure}{0.27\textwidth}
         \centering
         \includegraphics[width=\linewidth]{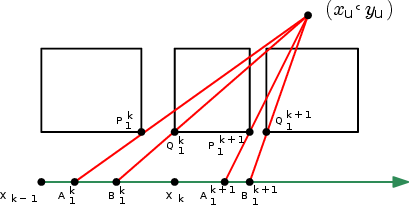}
         \caption{User-UAV link}
         \label{fig:2a}
     \end{subfigure}
     \begin{subfigure}{0.19\textwidth}
         \centering
         \includegraphics[width=\linewidth]{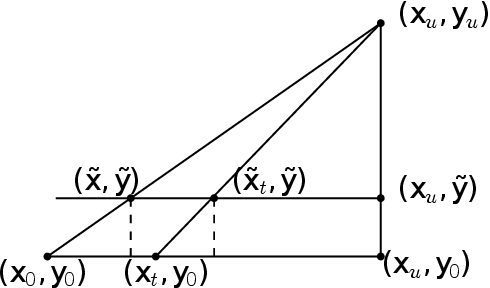}
         \caption{Proof of Lemma \ref{lem:p-los}}
         \label{fig:2b}
     \end{subfigure}
     \caption{User movement along X-axis}
     \label{fig:2}
    \vspace{-15pt}
\end{figure}

\noindent Let the user move with velocity $v$ along the positive X-axis, the buildings are placed according to the PPP with parameter $\lambda$, and the building heights are drawn from a Rayleigh distribution with parameter $\sigma$ (Fig. \ref{fig:2a}). Let $\hat{g}_t=(x_t,y_t)$ be the position of the user at time $t\in [0,T]$. Our motivation is to track the projected LoS path at any time $t$ to calculate the LoS probability of the user with the UAV. We know from (\ref{eq:p_los}) that the LoS probability depends on the coordinates of the first intersecting sides of the building and it alters over time due to the mobility of the user. With time, the sides are alternatively changing to a side parallel to the X-axis and then parallel to the Y-axis and vice versa. 

Let us first assume that the projected LoS path does not alter the first intersecting side parallel to the X-axis during the time interval $[0,T_\alpha]$. Under this assumption, the following lemma \ref{lem:p-los} gives an expression of the LoS probability at any time $t\in [0,T_\alpha]$ and also a closed form expression of the expected LoS time for the user in this time interval.
\begin{lemma}
    Let the first intersecting side of a projected LoS link be parallel to the X-axis, which does not change over the time interval $[0,T_\alpha]$. The LoS probability of the user at any time $t \in [0,T_\alpha]$,  will be
\begin{align}
        P_{LoS}(\hat{g}_t,\hat{u}) = P_{LoS}(\hat{g}_0,\hat{u}) \times \exp (Avt)\label{xpara}
\end{align}
    where $P_{LoS}(\hat{g}_0,\hat{u})$ is the initial LoS probability between the user and the UAV as in (\ref{eq:p_los}). The expected LoS time $E[T_{LoS(0,T_\alpha)}]$ of the user in this interval $[0,T_\alpha]$ will be,
\begin{align}
        E[T_{LoS(0,T_\alpha)}] = P_{LoS}
(\hat{g}_0,\hat{u})\times\frac{1}{Av}(\exp(AvT_\alpha) - 1)   
    \end{align}
    \label{lem:p-los}
    \vspace{-0.5cm}
\end{lemma}
$Proof$: Let the user be at $\hat{g}_0=(x_0,y_0)$ at $t=0$ and then start moving with velocity $v$ along the positive X-axis. At time $t \in [0,T_\alpha]$, let the new location of the user be $\hat{g}_t=(x_t,y_t)$. Thus, $x_t = x_0 +vt$ and $y_t = y_0$.
Now, let $(\tilde{x}_t,\tilde{y}_t)$ be the coordinates of the first point of intersection of the projected LoS link with the building after time $t \in [0,T_\alpha]$. So, using (\ref{eq:p_los}), the LoS probability of the user at time $t \in [0,T_\alpha]$ is
\begin{align}
     P_{LoS}(\hat{g_t},\hat{u}) =& \Big[1- exp\Big(-\frac{h_u^2}{2\sigma^2}\frac{|\tilde{x}_t-x_t|^2}{|x_u-x_t|^2}\Big)\Big]\notag\\\times&\exp\Big( A'|x_u - x_t| + B'|y_u - y_t| \Big)
     \label{eq:p_los1}
\end{align}
where 
\begin{align}
    A' = -\lambda\sqrt{\frac{\pi}{2}}\frac{\sigma}{h_u}(\operatorname{erf}(\frac{h_u}{\sqrt{2}\sigma}) - \operatorname{erf}(\frac{h_u}{\sqrt{2}\sigma} \frac{|\tilde{x}_t - x_t|}{|x_u - x_t|}))\nonumber \\
    B' = -\lambda\sqrt{\frac{\pi}{2}}\frac{\sigma}{h_u}(\operatorname{erf}(\frac{h_u}{\sqrt{2}\sigma}) - \operatorname{erf}(\frac{h_u}{\sqrt{2}\sigma}  \frac{|\tilde{y}_t - y_t|}{|y_u - y_t|})).\nonumber
\end{align}
Here, $\tilde{x}_t = \tilde{x} +\frac{|y_u-\tilde{y}|}{|y_u-y_0|}vt$ and $\tilde{y}_t = \tilde{y}$. Again, from Fig. \ref{fig:2b},

\begin{align}
    \frac{|\tilde{y}_t-y_t|}{|y_u-y_t|}=\frac{|\tilde{x}_t - x_t|}{|x_u-x_t|}=\frac{|\tilde{y}-y_0|}{|y_u-y_0|}=\frac{|\tilde{x}-x_0|}{|x_u-x_0|},
    \label{eq1}
\end{align}
which implies $P^0_{LoS}(\hat{g_t},\hat{u})$ at any time $t$, $0<t \leq T_\alpha$, is the same as $P_{LoS}^0$ at $t=0$ as obtained in (\ref{eq:p-los0}). This in turn implies that $A' = B'= A = B$ for all $t\in[0,T_\alpha]$. Hence the LoS probability $P_{LoS}(\hat{g_t},\hat{u})$ of the user at time $t \in [0,T_\alpha]$ given by equation (\ref{eq:p_los1}) can be rewritten as, 
\begin{equation*}
\vspace{-0.1cm}
    \begin{aligned}
         &= P^0_{LoS}(\hat{g_0},\hat{u})\exp\Big( A'|x_t - x_u| + B'|y_t - y_u|  \Big)\\
        &= P^0_{LoS}(\hat{g_0},\hat{u})\exp\Big( A|x_0 + vt - x_u| + B|y_0 - y_u|\Big) \\
        &=P^0_{LoS}(\hat{g_0},\hat{u})\exp\Big( A|x_0 - x_u| + B|y_0 - y_u|+Avt\Big) \\
        &= P_{LoS}(\hat{g}_0,\hat{u}) \times \exp(Avt)   \text{ (Using (\ref{eq:p_los}))}
    \end{aligned}
\end{equation*}
By integrating this expression with respect to $t$ in $[0,T_\alpha]$, we will get the expected LoS time of the user in $[0,T_\alpha]$ as,
\begin{align*}
\vspace{-0.2cm}
    E[T_{LoS(0,T_\alpha)}] &= \int_{0}^{T_\alpha} P_{LoS}(\hat{g}_t,\hat{u}) dt\\
     &= \int_{0}^{T_\alpha} P_{LoS}(\hat{g}_0,\hat{u}) \times \exp(Avt) dt\\
     &= P_{LoS}(\hat{g}_0,\hat{u})\times\frac{1}{Av}(\exp(AvT_\alpha) - 1)\qed
\end{align*}

Now we consider the case where the LoS link is first blocked by the side of a building block parallel to the Y axis, and it does not change during the time interval $[0,T_\beta]$. The following lemma \ref{lem:p_los_y} will give the expression for the expected LoS time in the time interval $[0,T_\beta]$.
\begin{lemma}
    The expected LoS time $E[T_{LoS(0,T_\beta)}]$ of the user in the time interval $[0,T_\beta]$ is
\begin{align}
=& \int_0^{T_\beta} P_{LoS}(\hat{g}_t,\hat{u})dt\notag\\
    \approx& \frac{T_\beta}{6}\Bigg[P_{LoS}(\hat{g}_0,\hat{u})+4P_{LoS}(\hat{g}_{\frac{T_\beta}{2}},\hat{u})+P_{LoS}(\hat{g}_{T_\beta},\hat{u})\Bigg]  \label{lem2}       
\end{align}
\label{lem:p_los_y}
\vspace{-0.2cm}
\end{lemma}
$Proof$: The proof goes along similar lines as that of Lemma \ref{lem:p-los}. However, in this case, for any $t \in [0,T_\beta]$, we get $x_t = x_0 +vt$, $y_t = y_0$, $\tilde{x}_t = \tilde{x}$, $\tilde{y}_t = \tilde{y} - \frac{|y_u-y_0|}{|x_u-x_0|}vt$. Hence the LoS probability $P_{LoS}(\hat{g_t},\hat{u})$ of the user at time $t \in [0,T_\beta]$ given by equation (\ref{eq:p_los1}) can be rewritten as, 
{\small
\begin{align}
    P_{LoS}(\hat{g_t},\hat{u}) =& \Big[1- exp\Big(-\frac{h_u^2|\tilde{x} - x_0 - vt|^2}{2\sigma^2|x_u -x_0-vt |^2}\Big)\Big]\notag\\&\times\exp\Big( A''|x_u-x_0-vt| + B''|y_u-y_0| \Big),
\end{align}}
\vspace{-0.5cm}
{\small
\begin{align}
    A'' =& -\lambda\sqrt{\frac{\pi}{2}}\frac{\sigma}{h_u}\Bigg[\operatorname{erf}(\frac{h_u}{\sqrt{2}\sigma}) - \operatorname{erf}\Big(\frac{h_u}{\sqrt{2}\sigma} \frac{|\tilde{x} - x_0-vt|}{|x_u-x_0 -vt|}\Big)\Bigg]\notag\\
    B'' =& -\lambda\sqrt{\frac{\pi}{2}}\frac{\sigma}{h_u}\Bigg[\operatorname{erf}(\frac{h_u}{\sqrt{2}\sigma}) - \operatorname{erf}\Big(\frac{h_u}{\sqrt{2}\sigma}  \frac{|\tilde{y}-\frac{|y_u-y|}{|x_u-x|}vt - y_0}{|y_u - y_0|}\Big)\Bigg].\nonumber
\end{align}}
In this case, no similar closed form expression of $P_{LoS}(\hat{g_t},\hat{u})$ like (\ref{xpara}) is obtained as the ratio given by (\ref{eq1}) does not hold. It also implies that $A'' \neq B''$ for any value $x_0\leq x_t \leq x_0+vT_\beta$. So, to calculate the expected LoS time $E[T_{LoS(0,T_\beta)}] = \int_0^{T_\beta} P_{LoS}(\hat{g}_t,\hat{u})dt$ of the user in $[0,T_\beta]$, we use Simpson's $1/3$-Rule to get (\ref{lem2}).\qed

So far we have computed the expected time under LoS for two scenarios, viz., when the LoS link is first blocked by the side of the building parallel to X and Y axes, respectively. However, as the user moves along its path, these two events will happen alternatively, all over the considered time epoch $[0,T]$. Let  $A_1^1,B_1^1$; $A_1^2,B_1^2$; $\cdots$; $A_1^k,B_1^k$ be the user positions on the roads where the projected LoS link changes its first intersecting sides and $t_A^1,t_B^1$; $t_A^2,t_B^2$; $\cdots$; $t_A^k,t_B^k$  be their respective time points. Our main objective would thus be to consider the sum of expected LoS time, for all such intervals. For each such interval, we can compute the expected LoS time using Lemma \ref{lem:p-los} and Lemma \ref{lem:p_los_y} respectively.

Consider $\hat{x}_1$, $\hat{x}_2$, $\cdots$, $\hat{x}_k$ are the x-coordinates of the points drawn from the MPLP. Let $P_1^1,Q_1^1$; $P_1^2,Q_1^2$; $\cdots$; $P_1^k,Q_1^k$ be the left and right corner points of the buildings.
 
From Fig. \ref{fig:2a}, 
\begin{eqnarray}
Q^k_1=\hat{x}_k,\;\text{and}\; P^k_1=\lambda|\hat{x}_k\mu_b+\hat{x}_{k-1}\mu_s|,\\   
\frac{y_{u}}{x_{u} - A_1^k} = \frac{y_{u}-w}{x_{u}-P_1^k},\;\text{and}\;
\frac{y_{u}}{x_{u} - B_1^k} = \frac{y_{u}-w}{x_{u}-Q^k_1}. 
\end{eqnarray}
By solving the above equations, we get
\begin{eqnarray}
A_1^k = \frac{y_uP_1^k - wx_u}{y_u-w}, \;\; B_1^k = \frac{y_{u}Q_1^k - wx_u}{y_u-w},
\label{eq:B_k}
\end{eqnarray}
where $w$ is the known street width.
So $t_A^k = \frac{A_1^k}{v}$ \& $t_B^k = \frac{B_1^k}{v}$. Now, we calculate $E[T_{LoS(0,T)}]$ for the given time interval $[0,T]$ in the following Theorem \ref{the:1}.
\begin{theorem}
    Let there be $\ell$ event points in between the time interval $[0,T]$. So, the total LoS time of the user will be,
    \begin{align}
     E[T^{\ell}_{LoS(0,T)}] = E[T_{LoS(0,t_A^1)}]+\sum_{i=1}^{\ell} E[T_{LoS(t_A^i,t_B^i)}] \notag\\+ \sum_{i=1}^{\ell-1}E[T_{LoS(t_B^i,t_A^{i+1})}]+E[T_{LoS(t_B^{\ell},T)}]
     \label{ek_t}
    \end{align}\label{the:1}\vspace{-10pt}
\end{theorem}
$Proof$: Let the user cross $\ell$ intersections during $[0,T]$. So, there will be $\ell$ pairs of $A_1^i$ and $B_1^i$ in $[0,T]$. At each $A_1^i$ and $B_1^i$, the projected LoS link will alter. Hence,
\footnotesize{
\begin{align*}
E[T^{\ell}_{LoS(0,T)}] &=\int_0^TP_{LoS}(\hat{g_t},\hat{u})dt\\
    =\int_0^{t_A^1}  P_{LoS}&(\hat{g_t},\hat{u})dt + \sum_{i=1}^{\ell}\int_{t_A^i}^{t_B^i}P_{LoS}(\hat{g_t},\hat{u})dt+\cdots \\&+ \sum_{i=1}^{\ell-1}\int_{t_B^i}^{t_A^{i+1}}P_{LoS}(\hat{g_t},\hat{u})dt+\int_{t_B^{\ell}}^TP_{LoS}(\hat{g_t},\hat{u})dt\\
    = E[T&_{LoS(0,t_A^1)}]+\sum_{i=1}^{\ell} E[T_{LoS(t_A^i,t_B^i)}] \notag\\+& \sum_{i=1}^{\ell-1}E[T_{LoS(t_B^i,t_A^{i+1})}]+E[T_{LoS(t_B^{\ell},T)}]\qed
\end{align*}} 
\normalsize
\begin{figure*}
\centering
\begin{subfigure}[b]{0.22\textwidth}
\centering
\includegraphics[width=\linewidth]{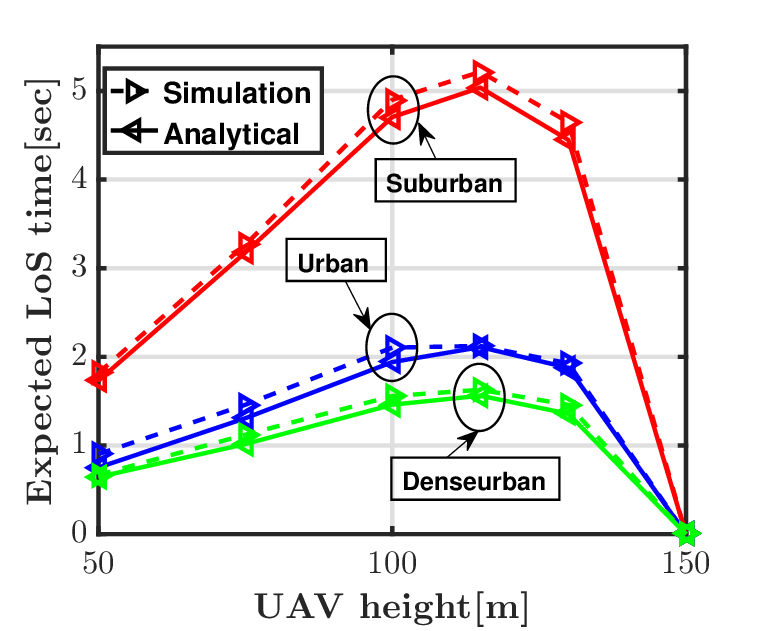}
\caption{}
\label{fig:graph1}
\end{subfigure}
\begin{subfigure}[b]{0.22\textwidth}  
\includegraphics[width=\linewidth]{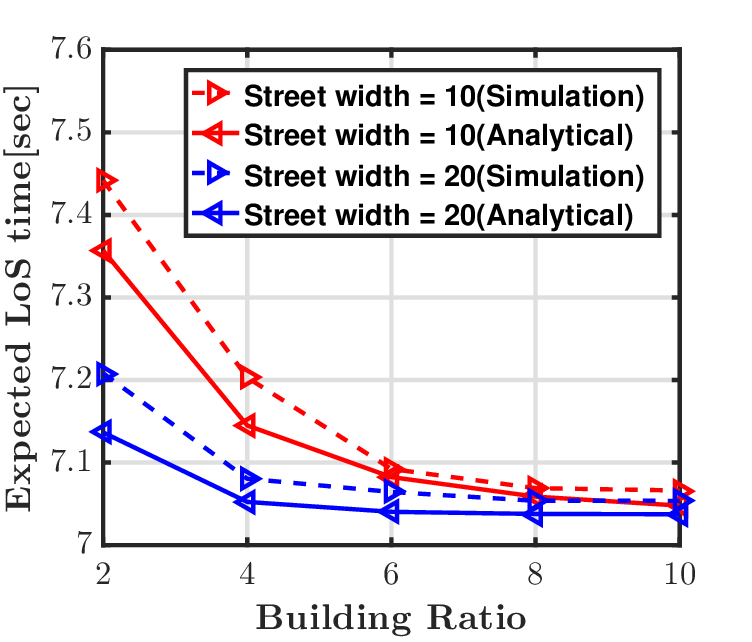}
\caption{}
\label{fig:graph2}
\end{subfigure}
\begin{subfigure}[b]{0.22\textwidth}  
\includegraphics[width=\linewidth]{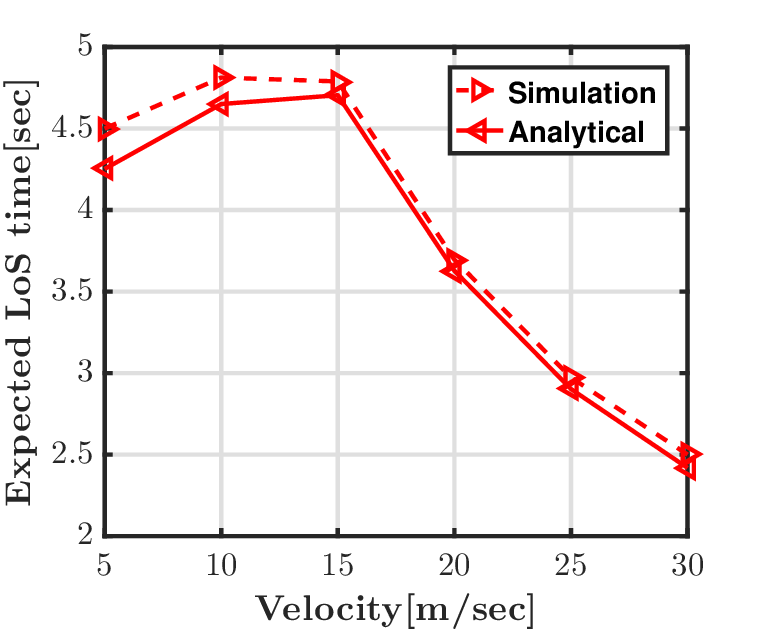}
\caption{}
\label{fig:graph3}
\end{subfigure}
\begin{subfigure}[b]{0.22\textwidth}
\includegraphics[width=\linewidth]{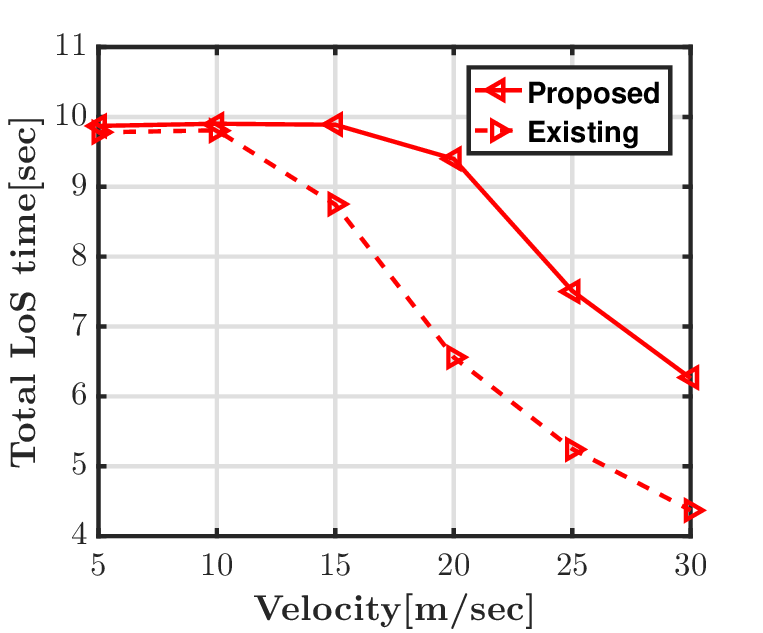}
\caption{}
\label{fig:graph4}
\end{subfigure}
\caption{(a) UAV heights vs expected LoS time, (b) Building ratio vs expected LoS time, (c) Velocity vs expected LoS time, (d) Time under LoS for existing and proposed schemes}
\vspace{-0.5cm}
\end{figure*}
The number of event points are drawn from PPP and hence $\ell$ is not fixed in $[0,T]$. In the following theorem we first calculate the probability of the highest number of events that can occur in $[0, T]$ and then compute the expected LoS time in $[0,T]$.
\begin{theorem}
    If $N$ is the highest number of events that occur in $[0, T]$ with probability $(1-\epsilon)$, then the expected total LoS time of the user will be,
\begin{align}
E[T_{LoS}]=\frac{\sum_{\ell=0}^{N}E[T^{\ell}_{LoS}]\times \frac{(\lambda vT)^{\ell} e^{-\lambda vT}}{\ell !}}{\sum_{\ell=0}^{N} \frac{(\lambda vT)^{\ell} e^{-\lambda vT}}{\ell !}}
\label{exp}
\end{align}
\label{main}
\vspace{-0.25cm}
\end{theorem}
$Proof:$ Let the user be at $\hat{g}_0$ and move with velocity $v$ along the X-axis in $[0,T]$.  Let $\mathcal{N}$ be the random variable that counts the number of events that occur in $[0,T]$. Clearly it follows a Poisson distribution with parameter $\lambda vT$. So, the highest number of events $N$ that can occur in $[0,T]$ with probability $1-\epsilon$ will be, 
\begin{align}
    P(\mathcal{N}\leq N) \geq 1 - \epsilon \implies P(\mathcal{N}\geq N) \leq \epsilon
\end{align}

Therefore, the value of $N$ will be,
\begin{align}
 N = \min \big\{ n \in \mathbb{N} : \sum_{j=0}^{n} \frac{\mu^{j} e^{-\mu}}{j!} \geq 1 - \epsilon \big\}   
\end{align}
where $\mu = \lambda vT$.

Now for each $\ell$, $0 \leq \ell \leq N$, $E[T^{\ell}_{LoS(0,T)}]$, the expected LoS time of the user in $[0,T]$, can be obtained by (\ref{ek_t}). The total expected LoS time $E[T_{LoS(0,T)}]$ can be calculated using the weighted average for all $\ell$ where each weight $w_{\ell}$ denotes the probability that $\ell$  number of events occur in $[0,T]$. So,
\begin{align*}
 E[T_{LoS(0,T)}] = \frac{\sum_{\ell=0}^Nw_{\ell}E[T^{\ell}_{LoS(0,T)}]}{\sum_{\ell=0}^Nw_{\ell}}   
\end{align*}
where $w_{\ell} =\frac{(\lambda vT)^{\ell} e^{-\lambda vT}}{\ell !}$.\qed

\subsection{UAV association}\label{uavasso}
\noindent Using the expected LoS time computed in Theorem \ref{main}, we derive a greedy user association algorithm that associates a user with the UAV that provides the maximum expected LoS time to it. Depending on the initial transmission distance and the user velocity, a user may not remain inside the UAV coverage area for the entire interval $[0,T]$. If we consider a UAV at height $h_u$, with a maximum 3D transmission distance of $d_{3D}$, the 2D coverage radius of the UAV becomes $\sqrt{d_{3D}^2-h_u^2}$. Obviously, the time $T'$ a user will spend within this coverage radius of the UAV depends on $v$, which might be less than $T$. Let $T_{min}$ be the minimum of $T'$ and $T$. Thus we shall calculate the expected LoS time of the user associated with the UAV under consideration during the time interval $[0,T_{min}]$. We compare our derived association policy with existing approaches like \cite{10499959,10064007,10198895} that associate a user to the nearest UAV providing the LoS by assuming user position to be static in $[0,T]$.

\section{Simulation Results}\label{simu}
\noindent In this section, we evaluate the numerical results for the expressions derived throughout the paper and validate the same using Monte Carlo simulations. We start by studying the correctness of our model by executing it in a variety of environments. 
Finally, we compare the proposed user association approach with an existing one \cite{10499959, 10064007,10198895}, and show the superiority in terms of time under LoS.
We consider three distinct grids, namely suburban, urban, and dense urban, with parameters as provided in\cite{6863654}, to generate our environment. The parameters are given in Table \ref{tab:params}.
\begin{table}[ht!]
\centering
\begin{tabular}{|p{1.5cm}||p{1.5cm}|p{1.5cm}|p{1.5cm}|} 
\hline
Type & Mean building height & Mean building width & Mean street width\\ 
\hline
Suburban & 10m & 37m& 10m \\
Urban & 19m& 45m& 13m\\
Dense urban & 25m& 60m& 20m\\
\hline
\end{tabular}\vspace{5pt}
\vspace{-0.2cm}
\caption{Grid parameters}\label{tab:params}
\vspace{-0.3cm}
\end{table}

We consider a $400\times 400$ $m^2$ region and create a grid by drawing points from an MPLP, whose parameters are drawn from Table \ref{tab:params}. The heights of the buildings are assumed to be drawn from a Rayleigh distribution with parameter $\sigma=8$. The user is initially at the origin (which can be an intersection or a road), and place a random UAV in the first or second quadrants (making use of symmetry). The user is moving with a velocity $v$ along the positive X-axis. We use Theorem \ref{main} to compute the expected time under LoS while considering the position of the UAV. To verify the accuracy of the results obtained, we check LoS along the user trajectory using a line sweep algorithm to compute the actual LoS time. This algorithm considers a 2D triangular area enclosed by the initial and final points of the ground user, and the UAV position. We consider the environment to be known, and efficiently compute the actual LoS time by using a classical line sweep algorithm.

This process is repeated $10000$ times and the average result is reported.

In Fig. \ref{fig:graph1} we compare the analytical result with the actual time under LoS for different heights of the UAV. The initial distance between the user and the UAV is kept constant, and the user moves with velocity $15$m/s along the X-axis. Increasing the height of the UAV will initially increase the expected LoS time of the user, upto a certain height determined by the coverage radius of the UAV. Thereafter, the coverage radius criterion nullifies any gain that comes with UAV height elevation. The correctness of the proposed analysis is validated by the close fit of the analytical results with the experimental ones. As is expected, the time under LoS decreases as we move from suburban towards urban, and dense urban grids.

In Fig. \ref{fig:graph2} we compare the analytical result with the simulation by varying the building ratios (ratio of building width to street width) for street width $10$m and $20$m. Here can see both analytically and by simulation that increasing the building ratios decrease the expected LoS time, as expected. Moreover, for street width $10$m the expected LoS time is greater than that for the street width $20$ as the total area occupied by the buildings is much lesser in the former than the latter.

In Fig. \ref{fig:graph3} we compare the analytical result with the simulated result for different velocities of the user. For low velocities, a blocked user will take a long time (possibly greater than $T$) to get out of non-LoS, while a moderately mobile user will get out of non-LoS zone fast; hence the initial increase in expected LoS time. However, for higher velocities, the time under the UAV coverage drops fast as the user quickly goes out of range of the UAV, and hence the decreasing trend of the expected LoS time.

In Fig. \ref{fig:graph4}, we plot the total LoS time after user association by two approaches. The existing approach assigns the nearest UAV without considering user mobility, while ours prioritizes the expected LoS time considering user mobility. We fix the UAV at a height of $100$m, and consider an urban environment, and plot the results by varying $v$. We see that at the lower velocities, the two approaches give similar results, while at higher velocities the proposed approach outperforms the existing approach by a considerable margin.

\section{Conclusion}\label{concl}
\noindent In this paper, we considered the often overlooked problem of time under LoS connectivity for a moving user served by UAVs. We use tools from stochastic geometry and fundamental probability to derive an analytical expression for the expected time under LoS for a mobile user on the ground in an urban scenario. We validate our proposed model using Monte Carlo simulations and demonstrate the utility of this metric by using it to allocate links greedily, and show its superiority compared to an existing approach. In our future work, we will incorporate non-uniform velocity of a user, with changing directions in particular, even in the small time epoch.
\bibliography{references2}
\end{document}